# Unravelling brittle fracture statistics out of self-healing patterns forming during femtosecond laser exposure

Christos-Edward Athanasiou[1], Max-Olivier Hongler[2], Yves Bellouard[1]

[1]Galatea Lab, STI/IMT, Ecole Fédérale Polytechnique de Lausanne (EPFL), Rue de la Maladière 71b, Neuchâtel, CH-2002, Switzerland

[2]STI/IMT, Ecole Polytechnique Fédérale de Lausanne (EPFL), Lausanne, Switzerland

**Abstract.** *Femtosecond laser written patterns at the surface of brittle materials may show a regenerative random transition from self-organized to disordered structures. Here, we show that this random intermittent behaviour carries relevant fracture statistics information, such as the so-called Weibull parameters. Furthermore, we draw a phenomenological analogy with idle and busy periods arising in queueing systems that we used to establish that these successive laser generated cycles are statistically independent. Based on this analogy and together with microscopic observations, we propose an experimental method bypassing the need for many specimens to build-up statistically relevant ensembles of fracture tests. This method is potentially generic as it may apply to a broad number of brittle materials.*

Under given femtosecond laser exposure conditions (1) (2), periodic patterns with sub-wavelength periodicity form in the bulk of various materials. Recently (3), we reported the occurrence of intermittent transitions between line bearing patterns and chaotic ones appearing at the surface of the material. To produce these patterns (shown in Fig. 1), we expose a specimen translated at a constant velocity under the focus of a femtosecond laser beam carrying 270 fs pulses at a rate of 800 kHz. These line patterns are written so that the laser affected zones intersect the surface. The transitions between organized parallel nanoplanes and randomly oriented cracks forming erratic patterns ('chaotic like'), occur abruptly and most interesting, alternates, displaying self-healing properties when apparent chaotic patterns reverse to highly organized ones (Fig. 1a).

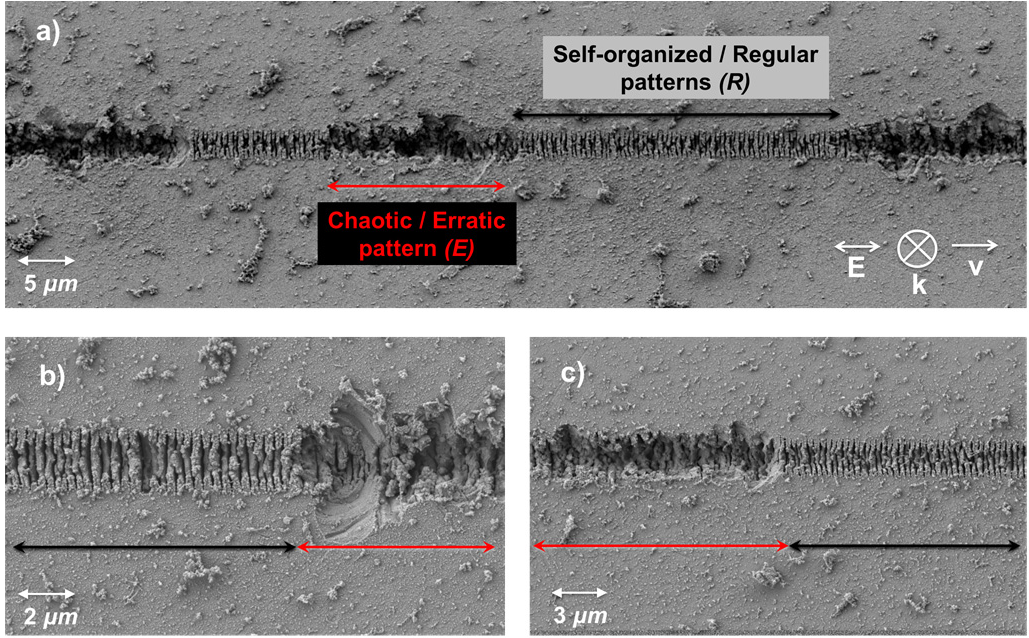

**Fig. 1:** *(a) Scanning Electron Microscope image (SEM) of laser processed line indicating the scanning direction (v), the polarization (E) and the pattern that is written on the surface of a fused silica substrate. (b) and (c) shows magnified view of some line patterns transitions between regular structures and chaotic structures and vice versa. The black double-arrows indicate regular patterns while the red ones emphasize chaotic regimes. The pulse energy is 196 nJ. There are about 40 pulses overlapping each unit volume producing a net fluence of 3 J/mm². In all these experiments the numerical aperture is 0.4.*

In this paper, we explore how this phenomenon can be used to extract information about the fracture mechanics of the surface.

Here, we formulate the hypothesis that these events find their origin in the statistical nature of the fracture. Indeed, for brittle materials, one cannot define a precise elastic limit above which they rupture. Rather, their fracture behaviour is usually described by a law that defines the probability of failure at a given stress level, well captured by a Weibull distribution (4) as follows:

$$P(m, \sigma_N; \sigma) = 1 - \exp\left(-\gamma \left(\frac{\sigma}{\sigma_N}\right)^m\right) \quad (1)$$

The above function is defined by three parameters: the nominal stress $\sigma_N$, the exponential factor m, and $\gamma$, a geometrical parameter fixed by the experimental conditions (in our case, the final and the initial experimental volume tested). The two main parameters, $\sigma_N$ and m, are determined by fitting an experimental curve derived from the equation above.

In previous works, we have shown that self-organized patterns are associated with a significant amount of stress introduced in the material during laser exposure (5). Having in mind this observation, we formulate the general hypothesis that *each single nanoplane is equivalent to loading the material to a certain stress level, and in other words, form a single 'nano-fracture' test experiment*. Furthermore, we postulate that *this phenomenon is generic and applies to a broad number of brittle materials*.

The general hypothesis is based on two assumptions. First, each single 'nano-fracture' test is independent to the previous one; second, each nanoplane forming at a given and known fluence corresponds to a known stress loading level. The second point has been validated in reference (6). Let us now establish the validity of the first one by examining the statistical nature of the intermittency.

To do so, a number of lines are written at the surface of three brittle materials using a femtosecond laser with varying deposited and pulse energies following a similar approach reported in (3). Then, we measure the length of sections displaying chaotic and self-organized patterns. The results are shown in Fig. 2 where black domains represent self-organized regimes (R) and red ones, chaotic regimes (E), respectively. For all the materials tested, fused silica, α-quartz and sapphire, we observe that as the deposited energy increases, both the length and frequency of erratic patterns increase.

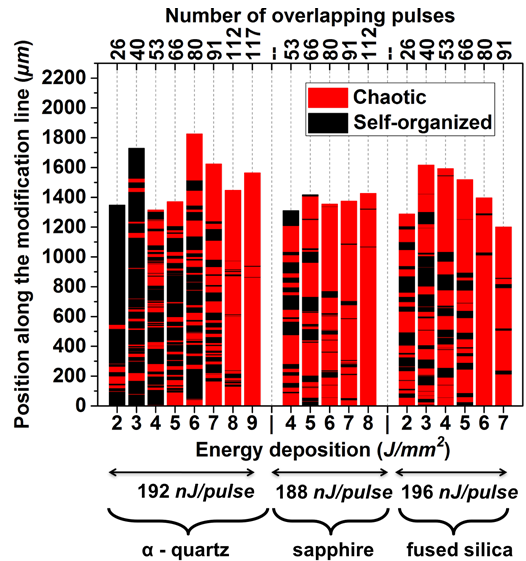

*Fig. 2: Overview of the transitions for different energy depositions and pulse energies for three different materials. Black regions represent the self-organized patterns and the red ones the chaotic patterns.*

In general, intermittency due to alternations between regular and erratic patterns may arise either in deterministic nonlinear dynamical systems (like Pomeau-Manneville (7)) performing transitions from periodic to chaotic evolutions or in stochastic storage systems, where an intrinsic randomness of the incoming and outgoing flows lead to an alternation between "busy" and "idle" periods in queuing theory (8). Here, we adopt the point of view and analogy of a *virtual queuing system* (QS).

Following Glansdorff-Prigogine theory (9), the laser / glass ensemble forms an open system in which the deposited energy is steadily dissipated via several distinct physical mechanisms, one of them being precisely the self-generation of intermittent dissipative patterns. Let us now analyse these intermittency using queueing dynamics framework. A single, cyclic alternation between the length $E$ and $R$ defines itself a cycle with random length $C = E+R$. From the observation, one may naturally raise the question: 'are the successive cycles ($C$) statistically independent?'

Following this metaphor, part of the laser energy pulses are stored into the glass and then released via the pattern generation mechanism, in the same way a waiting room 'storing' incoming customers is purged by delivering a service. From the general QS theory, we know that the queue content alternates between two states, the so-called idle periods (IP) and busy periods (BP). We shall call I, the time interval duration of an IP while $\Theta$ of a BP. Both I and $\Theta$ are random variables which, as we shall establish now, are statistically independent, a fact that will actually be experimentally verified *a posteriori*.

From microscopic observations, we note that the spacing between nanoplanes appearing in the regular portion of the patterns is essentially independent to the energy delivered by the laser (3). Considering this spacing, we may now define a natural characteristic length scale and its related time scale $t_0 := w_L/v_0$, where $v_0$ is the translation velocity of the laser. Let us also introduce $\lambda \in [0, 1]$ to be the probability that a busy period starts (or equivalently, an idle period ends) in the next time slot $t_0$. Accordingly, $Prob\{I = k\} = \lambda^{(k-1)}(1 - \lambda)$ stands for the geometric probability law to observe an idle period of duration k. For this geometric probability law, the average idle period time is $E\{I\} = \lambda^{-1}$ and is variance reads $\sigma_a^2 = (1-\lambda)/\lambda^2$.

Accordingly, we assimilate the beginning of the busy period start with the arrival of a virtual customer in an abstract discrete time queuing system belonging to the Geo/G/1 dynamics. The notation Geo stands for the geometric law governing the arrivals and G specifies the general service time distributions (let us write $\mu^{-1}$ and $\sigma^2_s$, for the average and the variance, respectively). From (8), the first two moments of the busy period $\Theta$ of the Geo/G/1 queue are explicitly known and reads:

$$\begin{cases} E\{\Theta\} = \dfrac{1}{\mu(1-\rho)}, \\ \sigma_\Theta^{\,2} = E\{\Theta^2\} - [E\{\Theta\}]^2 = \dfrac{\sigma_s^{\,2}}{(1-\rho)^3} \end{cases} \qquad (2)$$

where $\rho = \lambda/\mu \in [0, 1]$ is the QS offered traffic. Note that for the abstract QS, all times appearing in Eq (2) are now expressed in the natural time units $t_0$. As for an alternative stochastic process in general and hence for the Geo/G/1, the $E\{I\} = \lambda^{-1}$, we have (10):

$$\rho = \frac{E\{\Theta\}}{E\{I\} + E\{\Theta\}} \qquad (3)$$

Eq. 3 implies that $\rho$ can be measured directly from the experimental patterns in Fig. 2. In addition keeping $\mu$ fixed but varying the traffic r (i.e. the delivered deposited energy), the first line of Eq. 2 implies:

$$\frac{E\{\Theta_1\}}{E\{\Theta_2\}} = \frac{1-\rho_2}{1-\rho_1} \qquad (4)$$

where $\Theta_k$ and $\rho_k$ for $k = 1, 2$ stand for two different BP's resulting from two different traffic loads. Accordingly, whenever (up to experimental errors) the equality Eq. (4) is satisfied for a set of experiments, one can conclude that the hypotheses underlying our abstract Geo/G/1 picture hold. If this is the case, the further conclusions can be drawn:

a) a single $\mu$ exists and it characterizes the typical energy dissipation rate that can be associated to the R/E patterns generation.

b) Most important, the basic hypothesis underlying the QS theory is realized. *Since these conditions are fulfilled, the successive busy (BP) and idle periods (IP) as well as their alternation are statistically independent.*

The experimental validity for three different materials is expressed by Eq. 4 and is summarized in Table 1. This validates our abstract QS metaphor explicitly and it is reasonable to assume that each nanoplane forms a statistically independent fracture test experiment.

| α-Quartz | | | Sapphire | | | Fused silica | | |
|---|---|---|---|---|---|---|---|---|
| 192 nJ | | | 188 nJ | | | 196 nJ | | |
| 0.94 | 0.89 | 0.05 | 0.68 | 0.68 | 0.00 | 0.59 | 0.65 | 0.06 |
| 0.70 | 0.73 | 0.03 | 0.85 | 0.99 | 0.14 | 0.77 | 0.69 | 0.08 |
| 0.77 | 0.66 | 0.11 | <span style="color:red">0.62</span> | <span style="color:red">0.84</span> | <span style="color:red">0.22</span> | 0.24 | 0.25 | 0.01 |
| 0.24 | 0.15 | 0.09 | 0.68 | 0.60 | 0.08 | 0.50 | 0.60 | 0.10 |
| $E\{\Theta_i\}/E\{\Theta_j\}$ | $(1/\rho_i)/(1/\rho_j)$ | Absolute difference | <span style="color:red">0.55</span> | <span style="color:red">0.21</span> | <span style="color:red">0.34</span> | $E\{\Theta_i\}/E\{\Theta_j\}$ | $(1/\rho_i)/(1/\rho_j)$ | Absolute difference |
| | | | 0.22 | 0.14 | 0.08 | | | |
| | | | $E\{\Theta_i\}/E\{\Theta_j\}$ | $(1/\rho_i)/(1/\rho_j)$ | Absolute difference | | | |

*Table 1: Summary of the Geo/G/1 queuing theory results for α-quartz, sapphire and fused silica. For the calculations, measured values of $E\{\Theta_i\}$ and $\rho_i$ are used to validate the left and the right part of eq. 4. When this values coincide, the system follows the dynamics of the Geo/G/1 QS. Values within 15% are represented in green colour while outliers are represented in red colour.*

Let us now examine the microscopic aspect of the intermittency, and more specifically the mechanism of crack formation, which clearly triggers the transition to a chaotic regime. Interestingly, the laser deposited energy and the stress in the material are related as shown in fig 3b.

In previous works, we have demonstrated that the formation of nanogratings is accompanied by net volume expansion (6), which generates significant amount of stress in the material. We also know, that an increasing deposited energy on the material corresponds to the linear increase of the stress (5) as shown in Fig. 3. The material becomes porous (11), expands (6), generating compressive stress around the laser affected zones. High stress concentration is found at the tip of these nanoplanes, where the stress gradient is the highest. This hypothesis is also supported by atomic force microscope measurements (12) indicating the interface of the laser affected and unaffected zones as prone to crack nucleation and formation.

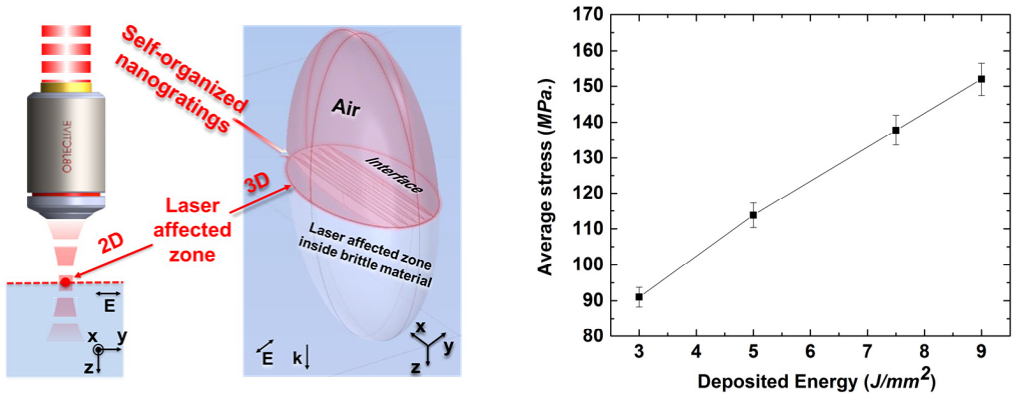

*Fig. 3: (a) Schematic illustration of the femtosecond laser ellipsoid which corresponds to the laser affected volume. The waist of the ellipsoidal is focused on the material-air interface, producing patterns on the surface of the material. (b) The average stress between the nanoplanes in fused silica for the case of 196 nJ per pulse at 800 kHz. The deposited energy is `tuned` by changing the translation velocity of the laser.*

The formation of individual elements of a pattern (nanoplane) is essentially a tensile test at the nanoscale. The maximum stress is found at the tip of the nanoplane, and in first approximation, is given by Inglis formula (13): $\sigma_{max}=\sigma_{AVG}(1+2a/b)$. Parameters $\sigma_{AVG}$, a, b are the average stress between the lamellae, and geometrical factors, respectively (see fig. 4b). Based on previous work (6), the average stress within nanoplanes as a function of the deposited energy can be precisely determined for each specific laser exposure conditions and is found to linearly increase with the amount of energy deposited in the material (fig. 4b).

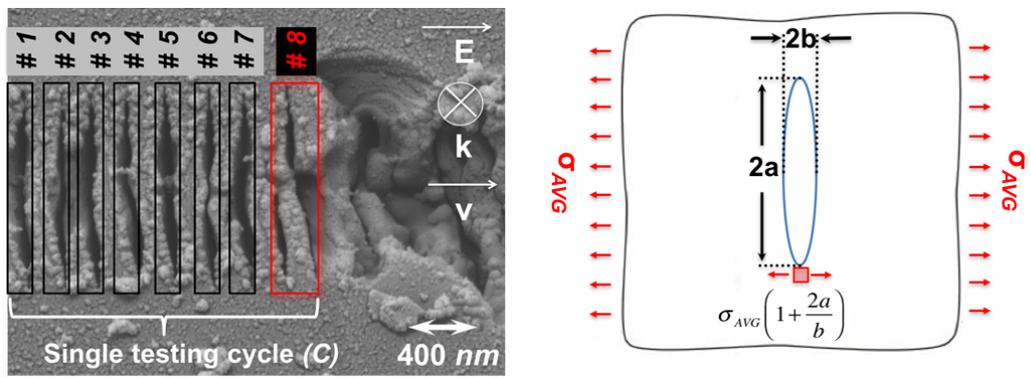

*Fig. 4: (a) Each element formation can be seen as an individual tensile-stress loading experiment. A testing cycle (C) is terminated a fracture occurs. (b) Illustration of the loading conditions for a single nanoplane formalized as an Inglis fracture problem (13).*

Each single experiment *per se*, can have two outcomes: either one regular nanoplane formation or the nucleation of a crack, respectively. If we assume an strict independence between loading events as discussed in the previous paragraphs, the number of experiments taking place before failure occurs, can be formulated mathematically by the geometric distribution $Prob\{r = n\} = p(1-p)^{n-1}$, where $r$ is the number of experiments performed in one experimental cycle (defined as C in our QS model) and $p$ is the probability of failure to happen.

In the brittle materials considered here, one cannot precisely define the elastic limit above which the material ruptures. The probability of failure under a given stress is commonly described by a Weibull statistical law (4). Combining the Weibull law with the average of geometric distribution, we get the following equation:

$$\underbrace{\ln\left(-\ln\left(1-\frac{1}{<r>}\right)\right)}_{y} = m\ln(\sigma) = \underbrace{m\ln\sigma}_{mx} - \underbrace{(m\ln\sigma_N + \ln\gamma)}_{\beta} \qquad (6)$$

The equation is defined by two parameters: the scale parameter which has the same dimensions as a stress, and the dimensionless factor m (also called Weibull modulus). For the case of fused silica, the average number of nanoplanes ($<r>$) is measured using the SEM images like the ones shown in fig 2. Parameter $\gamma$ is needed in order to take into account the different size of these stretched ellipsoidal nanoplanes as the energy per pulse increases.

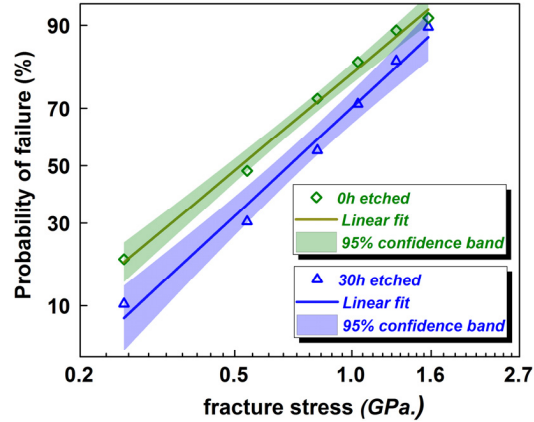

*Fig. 5: The graph represents the probability of fused silica to fail under a certain stress level. The two curves correspond to two different surface qualities. For both cases, the 95% confidence band is given by the equation $x \pm Z$ $x \sigma / sqrt(n)$ where Z is the confidence level, σ is the standard deviation and n the sample size. For the two cases, the relative error along the x-axis is 3% of the values and finds its origin in the evaluation of the average stress between the nanoplanes (see Fig. 4). The relative error along the y-axis is estimated to be 5% of the values and is related to the uncertainty of the start position of the transition towards a chaotic regime. For the sake of clarity, the error bars are not presented in the graphs.*

By linearly fitting the data points, a visual assessment of the validity of the initial hypothesis – "nanoplanes form independent tensile tests" – is provided. As shown in fig. 5, the experimental data form a single straight line in accordance with eq. 6 with convincingly high levels of confidence. The maximum stress for our mechanical tests reaches 1.7 *GPa*. This value is consistent with previously reported ones in both silica fibres produced by fusion splicing (14) as well as for fused silica mechanical parts processed via femtosecond laser exposure and chemical etching (15) (16).

The theoretical strength value of bulk fused silica to fail is estimated in the order of 21 *GPa*. indicating that it fails due to the presence of surface flaws on its surface. It has been reported that chemical etching in hydrofluoric acid of silica does improve the mechanical resistance of glass surface (17) (16) (18). This result supports the validity of our hypothesis that every single nanoplane formation is essentially a fracture problem and essentially that the surface quality of the material is mechanically tested.

To summarize, we have investigated the intermittent behaviour when exposing the surface of brittle materials with femtosecond laser pulses. In particular, using queueing theory, we unravel striking similarities that we use to demonstrate that the formation of each nanoplane is independent of each other. This key finding in conjunction with our previous knowledge on the stress generation associated with the formation of these nanoplanes, leads us to conjecture that each single lamella formation is equivalent to a single '*loading*' *experiment*. Here, we prove that the intermittent behaviour finds its roots in the fracture mechanics of the material. From the number of nanoplanes formed before fracturing the material, one can extract important parameters of the glass' surface mechanics such as the Weibull parameters, commonly used for defining the probably of rupture for brittle materials.

Overall, this methodology offers a straightforward and contact-less method for extracting fracture mechanics data of surfaces at all size and with minimum amount of materials. It opens new opportunities for rapid diagnosis of surface strength, for instance for quality control of consumer electronics and other fields, a quality control test that is currently hard to implement; but also for further analysing the behaviour of brittle materials at small scales.

The authors acknowledge the financial support of the European Research Council (Galatea project, ERC-2012-StG-307442)